# Topologically protected entanglement switching around exceptional points

Zan Tang,* Tian Chen,*+ Xing Tang, and Xiangdong Zhang$

Key Laboratory of Advanced Optoelectronic Quantum Architecture and Measurements of Ministry of Education, Beijing Key Laboratory of Nanophotonics & Ultrafine Optoelectronic Systems, School of Physics, Beijing Institute of Technology, 100081, Beijing, China

*These authors contributed equally to this work. $+To whom correspondence should be addressed. E-mail: zhangxd@bit.edu.cn; chentian@bit.edu.cn

**The robust operation of quantum entanglement states are crucial for applications in quantum information, computing, and communications[1-3]. However, it has always been a great challenge to complete such a task because of decoherence and disorder. Here, we propose theoretically and demonstrate experimentally an effective scheme to realize robust operation of quantum entanglement states by designing quadruple degeneracy exceptional points. By encircling the exceptional points on two overlapping Riemann energy surfaces, we have realized a chiral switch for entangled states with high fidelity. Owing to the topological protection conferred by the Riemann surface structure, this switching of chirality exhibits strong robustness against perturbations in the encircling path. Furthermore, we have experimentally validated such a scheme on a quantum walk platform. Our work opens up a new way for the application of non-Hermitian physics in the field of quantum information.**

Quantum entanglement as the heart of quantum mechanics highlights the nonseparability and nonlocality, which has been created experimentally in various physical systems. However, it is susceptible to influences of environment, which often appears decoherence. How to perform robust entanglement operations is crucial for applications in quantum information[1-3]. Recent investigations have shown that the combination of topology and quantum states can bring hope to solve such a problem, including the topological quantum optics interface[4, 5], topological sources of quantum light [6-8], topologically protected two-photon quantum correlation[9-12], topologically robust transport of entangled photons [13, 14]. The problem is that the fidelities of entangled states become very low after these reported topologically protected operations. Although the transformation efficiency of entangled states can be improved by using inverse-design method[15-16], various parameters need to be designed for the transformation between different entangled states. Moreover, the signal also scatters to non-topologically protected channels, resulting in significant losses in the transformation of entangled states. Thus, how to realize robust entanglement operation with high fidelity is still unknown.

In this work, we provide topologically protected entanglement operations with high fidelity by designing quadruple degeneracy exceptional point (EP). The EP is a type of non-Hermitian degeneracy, and its research has attracted more and more attention[17-21]. This is because the abrupt nature of the phase transitions around or near the EP has been shown to lead to many intriguing phenomena, such as topological mode and energy transfers[22-28], laser mode selectivity[29-31], EP-enhanced mode splitting[32-44], loss-induced transparency[45, 46], unidirectional invisibility[47-48] and so on[49-52]. These phenomena have not only been explored in classical systems, but also they have been discussed in the quantum regime[53-55]. However, whether or how to achieve robust operations of entangled states around the EP has not yet been studied. As the first work on entanglement operations around the EP, our work opens up the exciting possibility of realizing robust entanglement operations with high fidelity in non-Hermitian systems.

**Theory of topological entanglement switching around degeneracy exceptional points**. In many previous studies[22-28], the dynamic encircling of the exceptional point was described using Hamiltonian operators. In the following, we provide another way to describe such a problem, that is, utilize non-Hermitian evolution operators based on non-Hermitian quantum walk (QW). As illustrated in Fig. 1a, two entangled particles (the red and gray spheres) as the input state $|\zeta_{in}\rangle$ are incident on an array composed of multiple operators $U_1, U_2, \cdots, U_N$. Here, the Hilbert space of each particle is $2\times1$ dimensional, and its state can be expressed using an orthogonal basis $|0\rangle=(1,0)^T$ and $|1\rangle=(0,1)^T$. The two entangled particles evolve along separate paths through the operator array, and the output entangled state can be expressed as $|\varsigma_{out}\rangle = U_N \cdot ... \cdot U_2 \cdot U_1 \cdot |\varsigma_{in}\rangle$. Each evolution step $U_i$ consists of the QW operator $I \otimes M_i$, as well as pre- and post-control operators $C_i$ and $C_i^{-1}$. The detailed expressions and derivations of these control operators are provided in Section 1 of Methods. Taking the first evolution step operator $U_1 = C_1(I \otimes M_1)C_1^{-1}$ as an example, the single step evolution process can be divided into three stages. First, both particles are acted upon by the operator $C_1$. Next, the red particle enters the identity matrix module $I$, while the gray particle enters the QW module $M_1$, where $M_1 = \psi(\varphi)R(\frac{\theta_1}{2})GSR(\theta_2)G^{-1}SR(\frac{\theta_1}{2})$ consists of multiple operators. Here $R(\theta) = \begin{pmatrix} \cos(\theta) & -\sin(\theta) \\ \sin(\theta) & \cos(\theta) \end{pmatrix}$ is the rotation operator, where $\theta_1/2$ and $\theta_2$ represent rotation angles. When $R(\theta)$ acts on the gray particle, it can make the output state a linear superposition state related to $|0\rangle$ and $|1\rangle$. $S = \begin{pmatrix} e^{ik} & 0 \\ 0 & e^{-ik} \end{pmatrix}$ is the conditional phase shift operator, which adds a phase shift of $e^{ik}$ for state $|0\rangle$, and the opposite phase shift of $e^{-ik}$ for state $|1\rangle$. The gain-loss operators are $G = \begin{pmatrix} e^{\gamma} & 0 \\ 0 & e^{-\gamma} \end{pmatrix}$ and $G^{-1} = \begin{pmatrix} e^{-\gamma} & 0 \\ 0 & e^{\gamma} \end{pmatrix}$, where $\gamma$ is the gain-loss strength. Under the action of $G$, the wave function with the state $|0\rangle$ ($|1\rangle$) is amplified (reduced). The effect of $G^{-1}$ is the opposite of $G$. The symmetry breaking operator is $\psi = \begin{pmatrix} \cos(\varphi) & i*\sin(\varphi) \\ i*\sin(\varphi) & \cos(\varphi) \end{pmatrix}$, which breaks parity-time symmetry when $\varphi \neq 0$. After the two particles have gone through $I \otimes M_1$, they are both acted upon by the operator $C_1^{-1}$. Afterwards, each evolution step operator $U_i(i \leq N)$ acting on the two particles follows a similar three-stage process, with the difference being that the parameters $\theta_1$ and $\varphi$ in the evolution change.

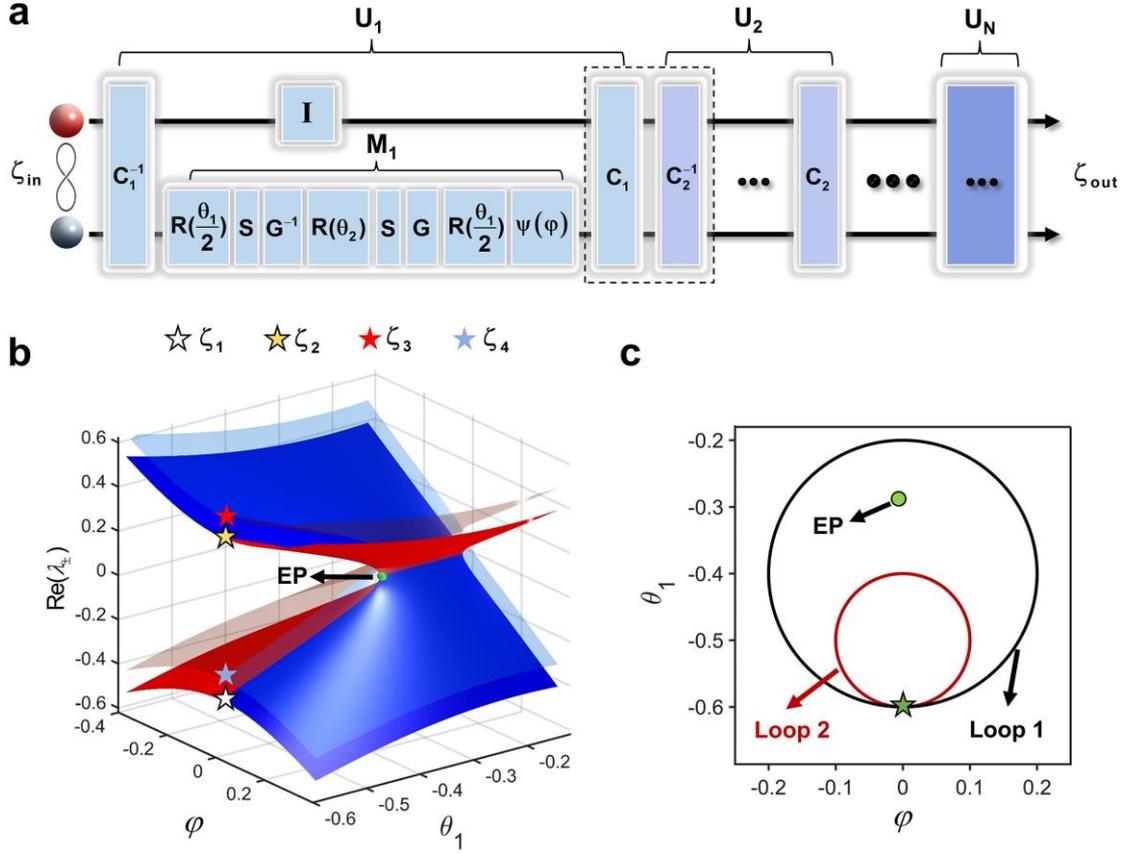

**Figure 1. The realization of the dynamic encircling of the exceptional point for entangled states. a** Schematic of the quantum walk evolution operator. **b** Riemann energy surface for evolution operator $U_i (i \leq N)$ with $\varphi$ and $\theta_1$, other parameters are $\gamma = 0.2, k = 0$ and $\theta_2 = \frac{\pi}{16}$. Four stars with different colors represent different initial states. A green ball indicates the position of EP. **c** Trajectories of the encircling path, where the black path Loop 1 surrounds the EP and the red path Loop 2 does not enclose or approach the EP.

Taking the parameters $\theta_1$ and $\varphi$ as variables, by solving the eigen-equation $U_i |\alpha_j\rangle = \eta |\alpha_j\rangle$ we obtain the eigenvalues $\eta_\pm = e^{-i\lambda_\pm}$, where $\lambda_\pm$ is the quasienergy of the evolution system, and $|\alpha_j\rangle (j=1,2,3,4)$ represent the four eigenstates of the evolution operator $U_i$. The details of these eigenvalues and eigenstates are provided in Section 2 of Methods. Figure 1b shows the real part of the quasienergy as a function of $\theta_1$ and $\varphi$. Four energy surfaces are divided into two groups where each group contains two degenerated Riemann energy surfaces. An isolated EP (green sphere in Fig. 1b) exists at the branch point of these surfaces.

The red or blue color indicates that the imaginary part of quasi-energy $\lambda_\pm$ is positive or negative, respectively. It is found that the evolution process described in Fig. 1a can exhibit behavior surrounding an EP by appropriately selecting the parameters $\theta_1$ and $\varphi$. When the parameter $(\varphi,\theta_1)=(0,-0.6)$ marked by the asterisk in Fig. 1b is chosen as the starting point, and $\varphi=0.2\times\cos(\pm\frac{2\pi}{N}n-\frac{\pi}{2})$ and $\theta_1=0.2\times\sin(\pm\frac{2\pi}{N}n-\frac{\pi}{2})-0.4$ ($N$ is the total number of step) at the $n$th step, the variation of parameters constitutes a loop (black Loop 1) as shown in Fig. 1c. The positive sign in the above equation corresponds to the entangled state evolving along the counter-clockwise path, while the negative sign corresponds to the clockwise path. Next, four Bell states $|\zeta_1\rangle=(|00\rangle+|11\rangle)/\sqrt{2}$, $|\zeta_2\rangle=(|00\rangle-|11\rangle)/\sqrt{2}$, $|\zeta_3\rangle=(|01\rangle+|10\rangle)/\sqrt{2}$ and $|\zeta_4\rangle=(|01\rangle-|10\rangle)/\sqrt{2}$ are taken as the input to the system, and the evolutions are studied. To make the state evolution approximately adiabatically, the number of total steps along Loop 1 is taken as $N=100$. In this way, the parameters $\theta_1$ and $\varphi$ change slowly. The theoretical density matrices of the four output states are shown in Fig. 2. Figs. 2e-2h correspond to the case where $\theta_1$ and $\varphi$ change clockwise, while Figs. 2i-2l correspond to the change counter-clockwise. For comparison, Figs. 2a-2d show the density matrices of the input states. The white, yellow, red and blue asterisks in Figs. 2a-2d corresponds to those labeled in Fig. 1b, which represent the input states $|\zeta_1\rangle$, $|\zeta_2\rangle$, $|\zeta_3\rangle$ and $|\zeta_4\rangle$, respectively.

As shown in Figs. 2e and 2f, when the input states $|\zeta_1\rangle$ and $|\zeta_2\rangle$ encircle the EP clockwise (CW), the evolved output results are both very close to the entangled state $|\zeta_2\rangle$. The calculated fidelities are as high as 98.3% and 96.4%, respectively. For comparison, if the input states $|\zeta_1\rangle$ and $|\zeta_2\rangle$ encircle the EP counter-clockwise (CCW), the output results are both very close to the entangled state $|\zeta_1\rangle$ with very high fidelities, see Figs. 2i and 2j. It indicates that encircling the EP enables asymmetric conversion between the entangled states $|\zeta_1\rangle$ and $|\zeta_2\rangle$.

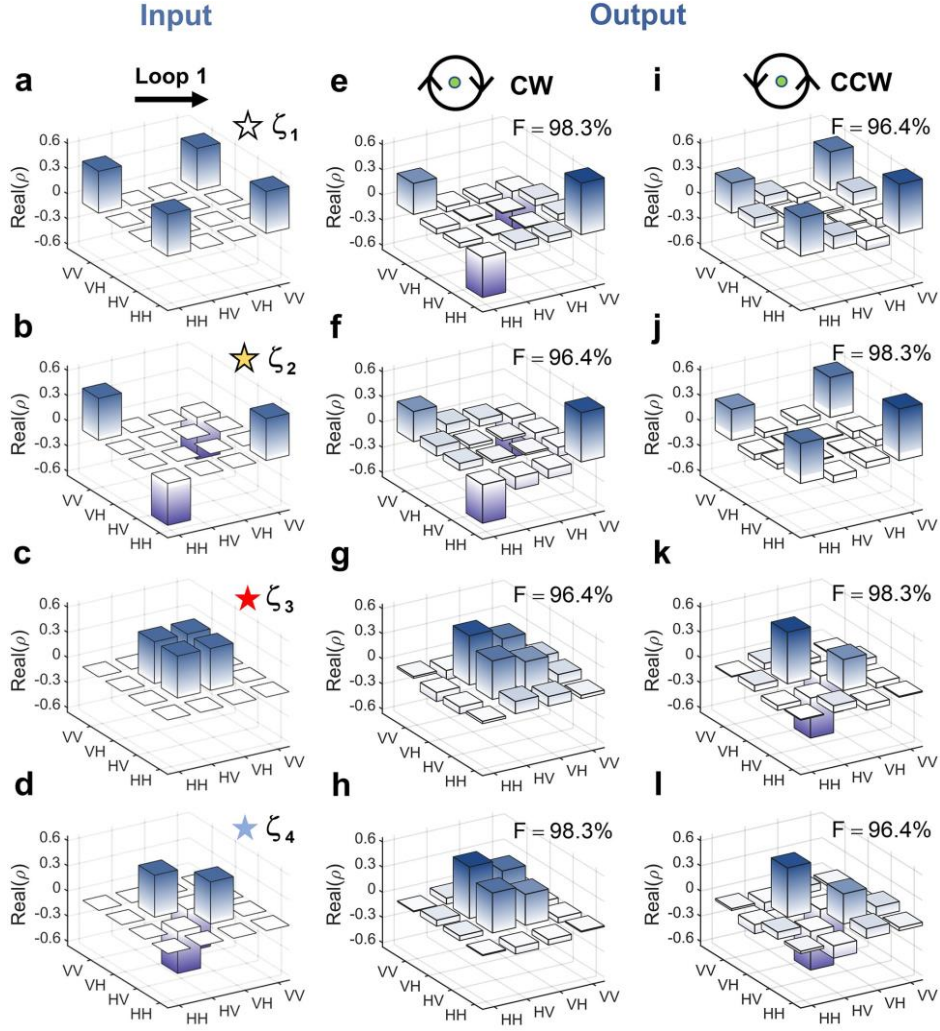

**Figure 2. Theoretical results for Loop 1 encircling the EP. a-d** Density matrices of the input states $|\zeta_j\rangle\,(j=1,2,3,4)$. **e-h** Density matrices of the output states after clockwise (CW) encircling, where **e** and **f** show the fidelity between the output states and $|\zeta_2\rangle$, and **g** and **h** show the fidelity between the output states and $|\zeta_3\rangle$. **i-l** Density matrices of the output states after counter-clockwise (CCW) encircling, where **I** and **j** show the fidelity between the output states and $|\zeta_1\rangle$, and **k** and **l** show the fidelity between the output states and $|\zeta_4\rangle$. At the top of **e** to **l**, the symbol F represents the fidelity between the output states and the ideal Bell states.

The origin for such asymmetric conversion is uncovered below. For the input state $|\zeta_1\rangle$, its real part of the energy is less than 0, which is described by the white asterisk in Fig. 1b. So when starting from $|\zeta_1\rangle$ and encircling the EP clockwise, the state experiences the evolution path on the red Riemann surface, i.e. experiencing the gain mode where the imaginary part of the quasienergy is positive. In this case, the input states adiabatically evolve on the Riemann surface, and change to $|\zeta_2\rangle$ after one cycle of parameter changes, as shown in Fig. 2e. While, when the input state changes to $|\zeta_2\rangle$ (its real part of the energy greater than 0, labeled as yellow asterisk in Fig. 1b), the evolution path encircling the EP clockwise is on the blue Riemann surface at the initial stage, i.e. the loss mode where the imaginary part of the quasienergy is negative. In this case, the tiny non-adiabatic coupling between the loss and gain modes of the non-Hermitian system induces non-adiabatic transitions, which breaks the adiabaticity. It results in the transition from the blue Riemann surface to the red one during the evolution, eventually return to itself after one cycle, as shown in Fig. 2f. Different evolution behaviors appear when the input states $|\zeta_1\rangle$ and $|\zeta_2\rangle$ encircle the EP counter-clockwise. For the input state $|\zeta_1\rangle$, the initial stage of the evolution paths are composed of the loss modes. So the non-adiabatic transitions occur during the evolution, causing the states to return to itself after one loop, as shown in Figs. 2i. While, when starting from $|\zeta_2\rangle$ and circling the EP counter-clockwise, the input state experiences the evolution path composed of the gain mode. Therefore, these input states evolve adiabatically on the Riemann sheet, and change to $|\zeta_1\rangle$ after one full period of parameters, see Fig. 2j. Detailed analysis of each step evolution have been provided in S1 of Supplementary Materials.

Similar results can also be found for $|\zeta_3\rangle$ and $|\zeta_4\rangle$. When the input states change to $|\zeta_3\rangle$ or $|\zeta_4\rangle$, and then encircles the EP clockwise, the output results are very close to the entangled state $|\zeta_3\rangle$ with high fidelities of 96.4% and 98.3%, see Figs. 2g and 2h. While, if the input state $|\zeta_3\rangle$ or $|\zeta_4\rangle$ encircles the EP counter-clockwise, the output results are both very close to the entangled state $|\zeta_4\rangle$, see Figs. 2k and 2l. These results mean that the asymmetric conversion between the entangled states $|\zeta_3\rangle$ and $|\zeta_4\rangle$ can also be realized through encircling the EP. The origin for such asymmetric conversion is also similar to that for $|\zeta_1\rangle$ and $|\zeta_2\rangle$.

The study above demonstrates that encircling the EP enables asymmetric conversion between the four entangled states, i.e., realize a chirality switch for entangled states. The output entangled state in the conversion is determined by the direction of circling the EP, and the conversion efficiency is very high. This phenomenon can be attributed to the consistency between the four eigenstates $|\alpha_j\rangle (j=1,2,3,4)$ of the evolution operator $U_1$ and the four input Bell states $|\zeta_j\rangle (j=1,2,3,4)$. To reveal this physical mechanism, the fidelities $F_j = \text{Tr}\sqrt{(\rho_{\xi_j})^{1/2} \rho_{\alpha_j} (\rho_{\xi_j})^{1/2}}$ $(j=1,2,3,4)$ between the four input Bell states and the eigenstates are calculated, where $\rho_{\xi_j} = |\xi_j\rangle\langle\xi_j|$ and $\rho_{\alpha_j} = |\alpha_j\rangle\langle\alpha_j|$. It is found that these fidelity values are all above 0.97, indicating the forms are very close. If the parameters are tuned to make the eigenstates $|\alpha_j\rangle (j=1,2,3,4)$ ideal Bell states, the output states will also be ideal Bell states. In addition, to achieve the above chiral switch, the evolution path of parameters cannot be far away from the EP. This chiral switch disappears if the evolution path of parameters are far away from the EP. For example, when the parameter values at the nth step are taken as: $\varphi = 0.1 \times \cos(\pm\frac{2\pi}{N}n - \frac{\pi}{2})$ and $\theta_1 = 0.1 \times \sin(\pm\frac{2\pi}{N}n - \frac{\pi}{2}) - 0.5$, they form a path not enclosing the EP but away from it, which is shown as the red Loop 2 in Fig. 1c. Our results of the Bell state conversion show that the chiral behavior disappears, which the detailed results has been provided in S2 of Supplementary Materials.

Furthermore, it is emphasized that the above manipulation processes for the entanglement states are topologically protected due to topological properties of EP. And more importantly, these phenomena can all be experimentally demonstrated. In the following, we discuss the experimental realization of the above theoretical scheme by constructing the non-Hermitian QW platform, and demonstrate the robustness of this chiral switch.

**Experimental realization of topological entanglement switching.** The constructed non-Hermitian QW platform is shown in Fig. 3, which contains three parts: state preparation (source), evolution process, and measurement. This corresponds to the theoretical scheme in Fig. 1a. In the state preparation, we first use 400 nm picosecond laser pulses to pump a 3 mm thick $\beta$-BaB$_2$O$_4$ (BBO) crystals, generating photon pairs at 800 nm through type-I parametric down conversion. These photon pairs are sent through interference filters to enhance their indistinguishability and coupled into single-mode fibers. The quantum states $|0\rangle$ and $|1\rangle$ of the two particles are encoded in the horizontal ($|H\rangle$) and vertical ($|V\rangle$) polarization states of the two photons, respectively. In the experiment, we choose the four maximally entangled Bell states $|\varsigma_{1,2,3,4}\rangle$ as the initial states. Since the operator $C_1^{-1}$ acting on the four Bell states $|\varsigma_{1,2,3,4}\rangle$ can yield product states that are easy to prepare accurately, we directly prepare the states $C_1^{-1}|\varsigma_{1,2,3,4}\rangle$ by rotating the angles of half-wave plates (HWP) and quarter-wave plates (QWPs), before sending them into the multi-step QW.

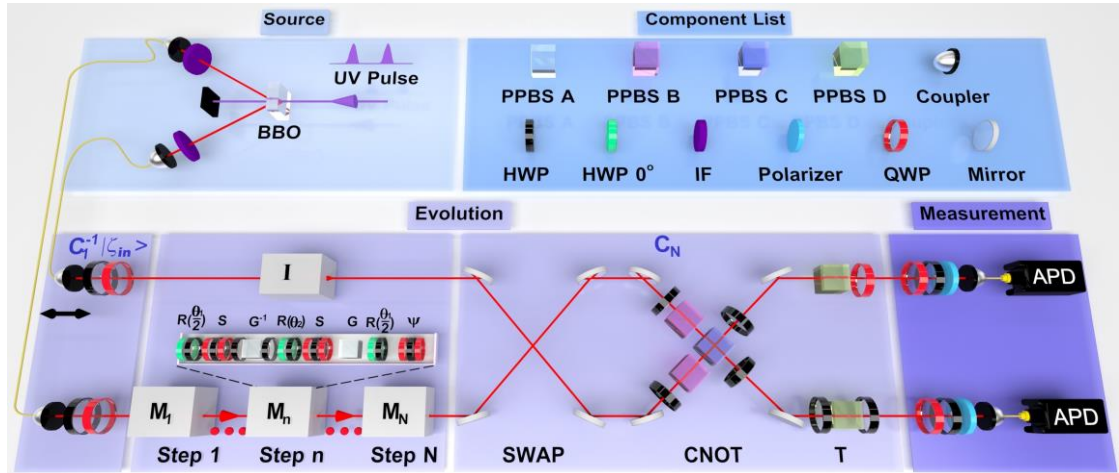

**Figure 3. Experimental setup. a** State preparation. **b** Implementation of the encircling evolution around the exceptional point. **c** Measurement.

After preparation, the photons are then sent into the multi-step QW $(I \otimes M_N) \cdots (I \otimes M_1)$. Compared with the previous theoretical design, the intermediate $C_i$ and $C_{i+1}^{-1}$ ($1 \leq i \leq N-1$) are omitted in the experiment, since $C_i C_{i+1}^{-1}$ is very close to the identity matrix for relatively large N, i.e., $C_i C_{i+1}^{-1} \approx I$, where the detailed analysis is provided in Section 3 of Methods. In the experiment, one photon propagates in free space while the other photon enters the QW $M_n$.

The operators in $M_n$ can all be implemented experimentally. The rotation operator $R(\theta)$ is implemented using a combination of a green HWP at $0°$ and a black HWP at $\theta$. Two QWPs and one HWP together implement the conditional phase shift operator $S$. For the gain-loss operator $G$, defining the relation $\gamma = 1/2\ln(l_1/l_2)$, we obtain an equivalent gain-loss operator $L = \begin{pmatrix} l_1 & 0 \\ 0 & l_2 \end{pmatrix}$ by equating small (large) loss to gain (loss), where $0 \leq l_1$ and $l_2 \leq 1$. Similarly, $G^{-1}$ can be implemented by the equivalent gain-loss operator $L^{-1}$. In the experiment, partially polarizing beam splitters (PPBS) are used to implement the operator $L$, while other loss operators $L^{-1}$ are implemented using a sandwich-type HWP-PPBS A-HWP optical device. Moreover, to implement the symmetry-breaking operator $\psi(\varphi)$, a combination of two QWPs and one HWP is placed at the end of each step. For different $M_n$, by changing the parameters $\theta_1$ and $\varphi$, the loops 1 and 2 described theoretically in Fig.1c are implemented experimentally. For the last operation $C_N$, we decompose it into the product of a SWAP gate, controlled-not (CNOT) gate, and the operator T. The SWAP gate can exchange the states of two quantum bits. In experiments, it can be implemented by exchanging the upper and lower photons using mirrors. The CNOT gate is implemented by Hong-Ou-Mandel interference using a combination of two PPBS B and one PPBS C. For the operator T, different combinations of HWPs, QWPs, and PPBS D are placed in the upper and lower paths to implement it. The detailed experimental implementation is provided in Section 4 of Methods.

After the two photons undergo the above evolution process, the output state is obtained through two-photon quantum state tomography. By using an apparatus consisting of QWPs, HWPs and polarizers, 16 measurement bases are constructed to perform projective measurements on the output state. With these projective measurement results, the quantum state tomography is completed and the density matrix of the output state is reconstructed.

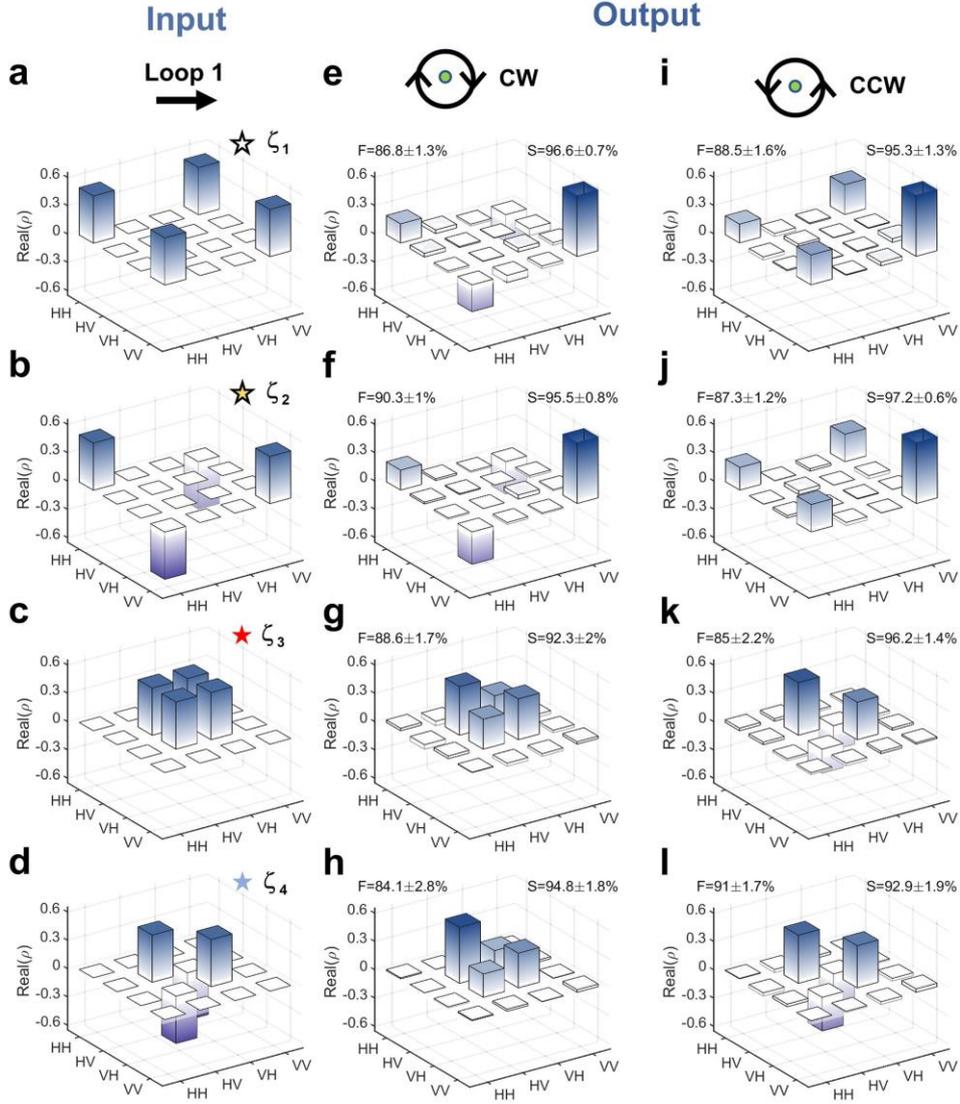

**Figure 4. Experimental results of the chiral entanglement switching with encircling an EP. a-d** Density matrices of the four different input Bell states $|\zeta_j\rangle$ $(j=1,2,3,4)$. **e-h** Experimental density matrices after clockwise encircling of the EP. **i-l** Experimental density matrices after counter-clockwise encircling of the EP. At the top of **e** to **l**, the symbol F represents the fidelity between the output states and the ideal Bell states. The symbol S represents the similarity $S[\rho_{th}, \rho_{ex}]$ between theoretical and experimental results.

The experimental results for the output states are shown in Fig. 4. In the experiment, a total number of QW steps $N=8$ is taken. In fact, the theoretical results shown in Fig.2 exhibit the case with 100-steps QW, which the parameters are unequally spaced. It is very difficult to realize experimentally with so many QW steps due to loss. Fortunately, it is found that good results can be obtained using fewer QW steps when the parameters $\theta_1$ and $\varphi$ are unequally spaced. This is because it is uneven for the matching degree between the output state and input state for each step evolution along Loops. The calculated results with $N=8$ are very close to those theoretical results with $N=100$, indicating that the adiabatic condition is also basically satisfied. Detailed analysis is provided in S3 of Supplementary Materials. When the initial states prepared in the experiment are $|\zeta_1\rangle$ and $|\zeta_2\rangle$, from Figs. 4e and 4f it is found that the final entangled states obtained are very close to $|\zeta_2\rangle$ when circling the EP clockwise; while from Figs. 4i and 4j, when circling the EP counterclockwise, the final entangled states obtained experimentally are very close to $|\zeta_1\rangle$. This demonstrates the chiral behavior of the entangled states $|\zeta_1\rangle$ and $|\zeta_2\rangle$ experimentally. In the same system, when the input states in the experiment are $|\zeta_3\rangle$ and $|\zeta_4\rangle$, both change leads to $|\zeta_3\rangle$ with encircling the EP clockwise (Figs. 4g and 4h); while leads to $|\zeta_4\rangle$ with encircling the EP counter-clockwise (Figs. 4k and 4l). These results are identical with those theoretical results shown in Fig. 2.

In the experiments, the fidelity $F=\text{Tr}\sqrt{\left(\rho_{\xi_j}\right)^{1/2}\rho_{ex}\left(\rho_{\xi_j}\right)^{1/2}}$ $(j=1,2,3,4)$ is calculated between the output state and the ideal entangled state, where $\rho_{ex}$ is the density matrix of the experimental output state and $\rho_{\xi_j}$ is one of the four Bell states. All fidelities reach 85% or above, indicating the output states are very close to the ideal entangled states. Because the total number of steps in the experiment is 8, there are some differences between the output states and the ideal entangled states, but it is sufficient to demonstrate the chiral switching of the entangled states. The similarity $S[\rho_{th},\rho_{ex}]=tr\sqrt{\rho_{th}^{1/2}\rho_{ex}\rho_{th}^{1/2}}$ between theoretical and experimental results is also calculated, where $\rho_{th}$ is the theoretical density matrix. It can be seen that the similarity for all cases is greater than 92%, indicating the excellent agreement between experiment and theory. This means that we have successfully experimentally demonstrated the chiral switch for the four Bell states. The inevitable loss of photon leads to the resource of error, and the related analysis in the experiment has been provided in S4 of Supplementary Materials. In addition, when choosing the red Loop 2 in Fig. 1c, the experimental results show that the chiral behavior disappears, which are also identical with theoretical results, see S5 of Supplementary Materials.

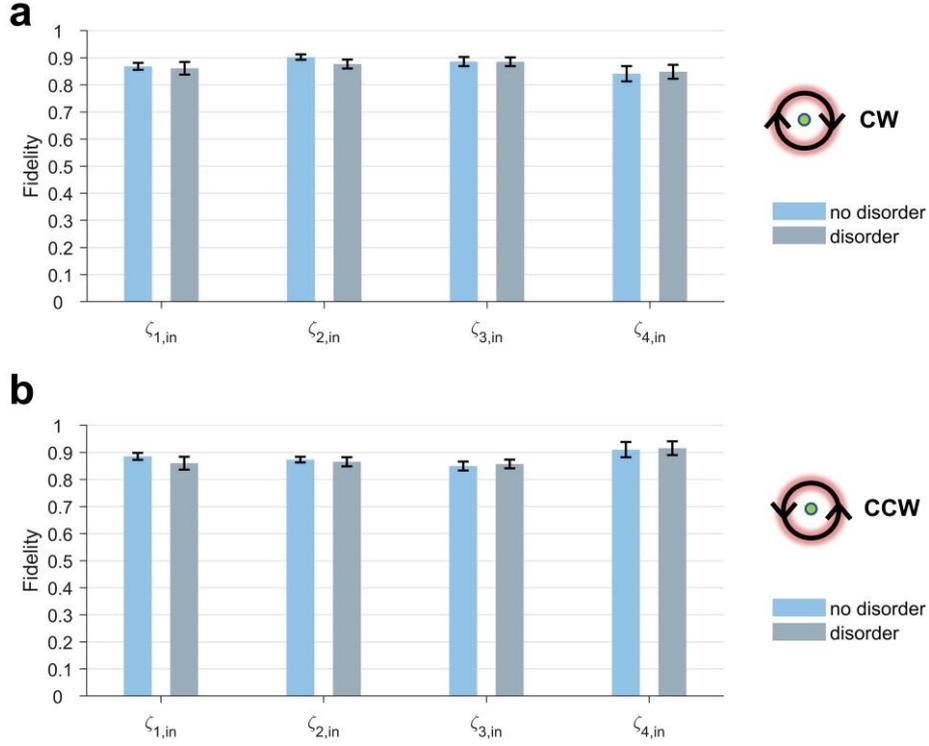

**Figure 5. Experimental results of fidelities.** In **a** and **b**, under the perturbation of disorder, the fidelity between the output states and the ideal entangled states for different input states. The label "CW" denotes the path circling the EP clockwise in the experiment, and "CCW" means counter-clockwise. The horizontal axis $|\varsigma_{j,in}\rangle (j=1,2,3,4)$ labels the four different input Bell states. The vertical axis represents the fidelity between the output state and the ideal output entangled state. The error bars without disorder represent ±1 s.d. estimated from Poisson photon counting statistics, and the error bars with disorder represent the standard deviation.

In order to verify the robustness of this switching behavior, the disorder is introduced into the encircled path, and to observe the variations in the output entangled states. In the experiment, the disorder is realized by adding small random angular deviations to the rotation angles of the waveplates, i.e., the parameters of path become $\theta_1 + \Delta\theta_1$ and $\varphi + \Delta\varphi$, where the disorder strengths $\Delta\theta$ and $\Delta\varphi$ are uniformly random chosen within the interval $(-0.025, 0.025)$ rad. Here, $\theta_1$ and $\varphi$ take the same values as those in the Fig. 4. Ten groups of perturbations are chosen, and the average results over these groups are shown in Fig. 5. The blue bars represent the fidelities between the output states and the ideal entangled states without disorder, and the gray bars represent the fidelities with disorder for comparison. Fig. 5a shows the cases of clockwise encircling of the EP. It can be seen that for the four different input Bell states, the fidelities between the output states and the ideal Bell states with disorder (gray bars) do not change much compared with the corresponding cases without disorder (blue bars), remaining at high values (above 0.85). Similar results are found for the cases of counterclockwise encircling of the EP in Fig. 5b. The fidelities with disorder also do not change much compared to the case without disorder. This means the chiral switching of the entangled states does indeed exhibit robustness against disorder in the path parameters.

**Discussion and conclusion.** The usual approach to achieving conversion of entanglement states is to precisely manipulate a two-qubit gate, and the conversion between different entangled states requires constructing different quantum gates. However, such an operation does not have topological protection characteristics, which is easily affected by environment and appears errors. In this work, we have provided effective scheme to realize robust operation of quantum entanglement states with high fidelity by designing quadruple degeneracy EPs. Because the designed Riemann energy surfaces with degeneracy EPs have the same eigenstates as the entangled states, asymmetric conversion between the entangled states can be realized by encircling the EP. Such manipulation for the entangled states is topologically protected due to the topological properties of the Riemann surface structure. Furthermore, the phenomena have been experimentally demonstrated by constructing the quantum walk platforms.

The above discussions focus on the case for encircling the EPs. Recent investigations have shown that chiral state transfers can appear without encircling the EP or near EP[27-28]. In fact, our designed topologically protected entanglement switching can also work without encircling the EP or near EP. The detailed discussions have been given in S6 of Supplementary Materials. This means that the phenomena we have revealed are easier to be implemented in various real systems, which is very beneficial for future quantum information, computing, and communications.

## Methods

**1. The details of pre- and post-control operators.**

In our discussion, the operators for the nth QW can be expressed as $M_n = \psi(\varphi) R(\frac{\theta_1}{2}) GSR(\theta_2) G^{-1} SR(\frac{\theta_1}{2})$. The terms $R(\frac{\theta_1}{2}) GSR(\theta_2) G^{-1} SR(\frac{\theta_1}{2})$ can be written by the $2\times 2$ identity matrix $\sigma_0$, and Pauli matrices $\sigma_x$, $\sigma_y$ and $\sigma_z$,

$$R(\frac{\theta_1}{2}) GSR(\theta_2) G^{-1} SR(\frac{\theta_1}{2}) = d_0\sigma_0 + d_x\sigma_x + id_y\sigma_y + id_0\sigma_z = \begin{pmatrix} d_0 + id_z & d_x + d_y \\ d_x - d_y & d_0 - id_z \end{pmatrix}. \quad (1)$$

The elements $d_0$, $d_x$, $d_y$ and $d_z$ are taken as $d_0 = \cos 2k \cos\theta_1 \cos\theta_2 - \cosh 2\gamma \sin\theta_1 \sin\theta_2$, $d_x = -\sinh 2\gamma \sin\theta_2$, $d_y = -\cos\theta_2 \sin\theta_1 \cos 2k - \cosh 2\gamma \cos\theta_1 \sin\theta_2$ and $d_z = \cos\theta_2 \sin 2k$, respectively. These elements satisfies the relation $d_0^2 - d_x^2 + d_y^2 + d_z^2 = 1$. When considering the operator $\psi(\varphi)$, the QW operator $M_n$ can be shown as

$$M_n = \begin{pmatrix} D_0 + iD_Z & D_X + D_Y \\ D_X - D_Y & D_0 - iD_Z \end{pmatrix} \quad (2)$$

with the operators $D_0$, $D_X$, $D_Y$ and $D_Z$ as

$$D_0 = \cos\varphi d_0 + i\sin\varphi d_x, D_X = \cos\varphi d_x + i\sin\varphi d_0, \\ D_Y = \cos\varphi d_y + \sin\varphi d_z, D_Z = \cos\varphi d_z - \sin\varphi d_y. \quad (3)$$

At the nth step, two photons undergo the evolution as $I \otimes M_n$, which can be described as

$$I \otimes M_n = \begin{pmatrix} D_0 + iD_Z & D_X + D_Y & 0 & 0 \\ D_X - D_Y & D_0 - iD_Z & 0 & 0 \\ 0 & 0 & D_0 + iD_Z & D_X + D_Y \\ 0 & 0 & D_X - D_Y & D_0 - iD_Z \end{pmatrix}. \quad (4)$$

For the above operator $I \otimes M_n$, it is obviously that the eigenstates are not the Bell states. Based on the studies about quantum state control encircling the EP, the efficient control among Bell states requires the eigenstates of system to be nearly Bell states. Therefore, to realize the efficient control of Bell states, we add the pre- and post-control operators $C_n$ and $C_n^{-1}$ to $I \otimes M_n$ in the theoretical design. In this way, the evolution operator at the *n*th step is

$U_n = C_n(I \otimes M_n)C_n^{-1}$. Due to the similarity transformation, the operators $U_n$ and $I \otimes M_n$ have the same eigenvalues as $\eta_\pm = D_0 \pm \sqrt{D_0^2 - 1}$. In the following, we provide the design of eigenstates of $U_n$.

When solving the eigen-equation $U_n$, we can always have the relation as,

$$U_n A = A \begin{pmatrix} \eta_- & 0 & 0 & 0 \\ 0 & \eta_- & 0 & 0 \\ 0 & 0 & \eta_+ & 0 \\ 0 & 0 & 0 & \eta_+ \end{pmatrix}. \tag{5}$$

Here, the column vectors in the matrix $A$ are composed of the eigenstates of $U_n$. For the operator $I \otimes M_n$, we can also have the eigen-equation as,

$$(I \otimes M_n) B = B \begin{pmatrix} \eta_- & 0 & 0 & 0 \\ 0 & \eta_- & 0 & 0 \\ 0 & 0 & \eta_+ & 0 \\ 0 & 0 & 0 & \eta_+ \end{pmatrix}, \tag{6}$$

where the column vectors in the matrix $B$ are composed of the eigenstates of $I \otimes M_n$. By combing Eq. (5) and (6), we can obtain the relation as,

$$U_n = AB^{-1}(I \otimes M_n) BA^{-1} = C_n(I \otimes M_n) C_n^{-1}. \tag{7}$$

The operator $C_n$ has the form as $C_n = AB^{-1}$. Through such similarity transformation, the eigenvalues of $I \otimes M_n$ and $U_n$ are the same, which means the same Riemann energy surfaces. The operator $U_n$ corresponds to the one step evolution for two photons, and can be expressed as

$$U_n = \begin{pmatrix} D_0 & 0 & D_Z & iD_X + iD_Y \\ 0 & D_0 & iD_X - iD_Y & -D_Z \\ D_Z & -iD_X - iD_Y & D_0 & 0 \\ iD_Y - iD_X & -D_Z & 0 & D_0 \end{pmatrix}. \tag{8}$$

2. **The construction of QW with Bell states as its eigenstates.**

By solving the eigen-equation, the eigenstates of $U_n$ can be obtained as $U_n|\alpha_j\rangle = \eta_\pm|\alpha_j\rangle (j=1,2,3,4)$, with eigenvalues $\eta_\pm = D_0 \pm \sqrt{D_0^2 - 1} = e^{-i\lambda_\pm}$. The explicit forms of $|\alpha_j\rangle (j=1,2,3,4)$ are

$$|\alpha_{1,2}\rangle = \frac{1}{\sqrt{2(\eta_\mp - D_0)}} \begin{pmatrix} iD_X + iD_Y \\ -D_Z \\ 0 \\ \eta_\mp - D_0 \end{pmatrix}, |\alpha_{3,4}\rangle = \frac{1}{\sqrt{2(\eta_\mp - D_0)}} \begin{pmatrix} D_Z \\ iD_X - iD_Y \\ \eta_\mp - D_0 \\ 0 \end{pmatrix}. \tag{9}$$

Considering the non-Hermitian system, the left eigenstates satisfy $U^\dagger|\beta\rangle = \eta^*|\beta\rangle$ with $|\beta_j\rangle (j=1,2,3,4)$ as

$$\langle\beta_{1,2}| = \frac{1}{\sqrt{2(\eta_\mp - D_0)}} \begin{pmatrix} -iD_X + iD_Y \\ D_Z \\ 0 \\ \eta_\mp - D_0 \end{pmatrix}^T, \langle\beta_{3,4}| = \frac{1}{\sqrt{2(\eta_\mp - D_0)}} \begin{pmatrix} -D_Z \\ -iD_X - iD_Y \\ \eta_\mp - D_0 \\ 0 \end{pmatrix}^T. \tag{10}$$

In our study, the parameters for the starting point of system are chosen as $(\varphi, \theta_1) = (0, -0.6)$. It is found that the eigenstates $|\alpha_{1,2,3,4}\rangle$ are very close to Bell states $|\zeta_{1,2,3,4}\rangle$. The fidelities between $|\alpha_{1,2,3,4}\rangle$ and $|\zeta_{1,2,3,4}\rangle$ are all larger than 97%. Therefore, the eigenstates of system with $(\varphi, \theta_1) = (0, -0.6)$ can be treated as Bell states. In this way, the efficient quantum control among bell states can be realized by encircling the EP.

3. **The simplification $C_{n+1}^{-1}C_n$ in the design of experiment**

When the evolution encircles the EP, the output state is obtained as:

$$|\zeta_{out}\rangle = U_N \cdots U_n \cdots U_2 U_1 |\zeta_{in}\rangle. \tag{11}$$

By replacing $U_n (1 \leq n \leq N)$ with its explicit form $C_n(I \otimes M_n)C_n^{-1}$, the above equation changes to:

$$|\zeta_{out}\rangle = C_N(I \otimes M_N)C_N^{-1} \cdots C_n(I \otimes M_n)C_n^{-1} \cdots C_1(I \otimes M_1)C_1^{-1}|\zeta_{in}\rangle. \tag{12}$$

In our study, the state changes slowly on the Riemann energy surface with $\varphi$ and $\theta_1$. For the adjacent $U_n$ and $U_{n+1}$, the parameters $\varphi$ and $\theta_1$ also change slowly. As mentioned in Section 1 of Methods, the operator $C_n$ has the form as $C_n = AB^{-1}$, where $A$ and $B$ are composed of the eigenstates of $U_n$ and $I \otimes M_n$, respectively. The slow changes of $\varphi$ and $\theta_1$ indicate that the expressions for $C_n$ and $C_{n+1}$ are nearly the same. So we can obtain the relation approximately as

$$C_{n+1}^{-1} C_n \approx 1. \tag{13}$$

For our discussions in the main text, we also numerically calculate the Eq. (13) and find it is always satisfied. In this way, the evolution shown in Eq. (12) can be simplified as:

$$|\zeta_{out}\rangle = C_N (I \otimes M_N) \cdots (I \otimes M_n) \cdots (I \otimes M_1) C_1^{-1} |\zeta_{in}\rangle. \tag{14}$$

The Eq. (14) means that we can achieve the circle of the EP following this simple evolution. In our experiment, the input states undergo the optical elements consisting of the evolution as Eq. (14), and change to different output states depending on the circle of the EP clockwise and counter-clockwise. Next, we show how to realize the operators in the optical platform.

### 4. Realizations of operators in the optical platform.

In the experiment, the combination of two QWPs and one HWP can realize any unitary operation of single polarization bits. Here, we design a specific combination of waveplates to achieve a more concise form of this evolution. The Jones matrices of the HWP and QWP are

$$HWP(\theta) = \begin{pmatrix} \cos(\theta) & \sin(\theta) \\ \sin(\theta) & -\cos(\theta) \end{pmatrix} \text{ and } QWP(\theta) = \frac{\sqrt{2}}{2} \begin{pmatrix} 1 - i\cos(\theta) & -i\sin(\theta) \\ -i\sin(\theta) & 1 + i\cos(\theta) \end{pmatrix}.$$

It is noted that the actual rotation angle in the experiment is $\frac{\theta}{2}$. In the following, we give the explicit realizations of these operators in the optical platform.

**i. Rotation operator $R$.**

$$R(\theta) = \begin{pmatrix} \cos(\theta) & -\sin(\theta) \\ \sin(\theta) & \cos(\theta) \end{pmatrix} = \begin{pmatrix} \cos(\theta) & \sin(\theta) \\ \sin(\theta) & -\cos(\theta) \end{pmatrix} \begin{pmatrix} 1 & 0 \\ 0 & -1 \end{pmatrix} = HWP(\theta) \cdot HWP(0), \tag{15}$$

The rotation operator $R$ can be achieved by the combination of two half-wave pieces with angles 0 and $\theta$ respectively. The symbol '·' represents the actions of operators from right to left.

**ii.    Conditional phase shift operator $S$.**

$$S = \begin{pmatrix} e^{ik} & 0 \\ 0 & e^{-ik} \end{pmatrix} = i\frac{\sqrt{2}}{2}\begin{pmatrix} 1 & -i \\ -i & 1 \end{pmatrix}\begin{pmatrix} \sin(k) & \cos(k) \\ \cos(k) & -\sin(k) \end{pmatrix}\frac{\sqrt{2}}{2}\begin{pmatrix} 1 & -i \\ -i & 1 \end{pmatrix}$$
$$= e^{i\frac{\pi}{2}} QWP(\frac{\pi}{2}) \cdot HWP(\frac{\pi}{2} - k) \cdot QWP(\frac{\pi}{2})$$
(16)

Therefore, the operator $S$ can be realized by combining two QWPs with an angle of $\frac{\pi}{2}$ and one HWP with an angle of $\frac{\pi}{2} - k$.

**iii.    The equivalent gain-loss operator $L$ and $L'$.**

The polarization-dependent loss operators $L = \begin{pmatrix} l_1 & 0 \\ 0 & l_2 \end{pmatrix}$ and $L' = \begin{pmatrix} l_2 & 0 \\ 0 & l_1 \end{pmatrix}$ are implemented using a partial polarization beam splitter (PPBS), an optical device with different transmittance $(t_H, t_V) = (l_1^2, l_2^2)$ for the horizontal and vertical polarizations of the incident light. The horizontal polarization in the experiment is set to be fully transmitted $(t_H = l_1^2 = 1)$, while the vertical polarization has a transmittance smaller than 1 $(t_V = l_2^2)$, realizing the polarization-controlled loss operators. The gain-loss strength $\gamma = 1/2\ln(l_1/l_2)$ is set to be 0.2, and the corresponding transmittance parameter of our customized PPBS is $(l_1^2, l_2^2) = (1, 0.45)$. Similarly, the operator $L'$ can be achieved by a PPBS with another type of transmittance $(t_H, t_V) = (l_2^2, l_1^2)$. In order to realize such a PPBS with a special polarization transmittance, we add two HWPs with a rotation angle of $\pi/4$ before and after injecting to the PPBS. So the operator $L'$ can be experimentally realized by a sandwich-type HWP-PPBS-HWP combination.

**iv.    The symmetry breaking operator $\psi$.**

$$\psi = \begin{pmatrix} \cos(\varphi) & i\sin(\varphi) \\ i\sin(\varphi) & \cos(\varphi) \end{pmatrix} = i\frac{1}{2}\begin{pmatrix} 1-i & 0 \\ 0 & 1+i \end{pmatrix}\begin{pmatrix} \cos(\varphi) & \sin(\varphi) \\ \sin(\varphi) & -\cos(\varphi) \end{pmatrix}\begin{pmatrix} 1-i & 0 \\ 0 & 1+i \end{pmatrix}$$
$$= e^{i\frac{\pi}{2}} QWP(0) \cdot HWP(\varphi) \cdot QWP(0)$$
(17)

The operator $\psi$ can be realized by combining two QWPs with an angle of 0 and one HWP with an angle of $\varphi$. We have provided the realization of $M_n$ in the optical platform, then the implementations of $C_1^{-1}|\varsigma_{in}\rangle$ and $C_N$ are given to complete the evolution encircling the EP.

v. **Realization of $C_1^{-1}|\varsigma_{in}\rangle$.**

In the experiment, the starting points of parameters are selected as $(\varphi,\theta_1)=(0,-0.6)$. The controlled operator $C_1^{-1}$ can be obtained with Eqs. (5)-(7).

$$C_1^{-1}=\begin{pmatrix} -i & 0 & 0 & 0 \\ 0 & 0 & 0 & 1 \\ 0 & 0.8071i & 0 & 0 \\ 0 & 0 & 1.2389 & 0 \end{pmatrix}. \tag{18}$$

When the input states are Bell states $|\varsigma_{1,2}\rangle$, the states undergoing the operator $C_1^{-1}$ are

$$C_1^{-1}|\varsigma_{1,2}\rangle=\frac{\sqrt{2}}{2}C_1^{-1}\begin{pmatrix} 1 \\ 0 \\ 0 \\ \pm 1 \end{pmatrix}=\frac{\sqrt{2}}{2}\begin{pmatrix} -i \\ \pm 1 \\ 0 \\ 0 \end{pmatrix}. \tag{19}$$

When the input states are Bell states $|\varsigma_{3,4}\rangle$, the states undergoing the operator $C_1^{-1}$ are

$$C_1^{-1}|\varsigma_{3,4}\rangle=\frac{\sqrt{2}}{2}C_1^{-1}\begin{pmatrix} 0 \\ 1 \\ \pm 1 \\ 0 \end{pmatrix}=\frac{\sqrt{2}}{2}\begin{pmatrix} 0 \\ 0 \\ 0.8071i \\ \pm 1.2389 \end{pmatrix}. \tag{20}$$

As shown by Eqs. (19)-(20), after going through the operator $C_1^{-1}$, the Bell states change to new product states. These product states can be realized through adding QWP and HWP to the two photons generated at the $\beta$-BaB$_2$O$_4$ (BBO) crystals.

vi. **Realization of $C_N$.**

According to the evolution encircling the EP (Eq. (14)), the final operator is $C_N$ with the parameters $(\varphi,\theta_1)=(0,-0.6)$. Through Eqs. (5)-(7), the explicit form of $C_N$ is

$$C_N=\begin{pmatrix} i & 0 & 0 & 0 \\ 0 & 0 & -1.2389i & 0 \\ 0 & 0 & 0 & 0.8071 \\ 0 & 1 & 0 & 0 \end{pmatrix}. \tag{21}$$

This operator can be realized by the combination of the SWAP, CNOT and T operators. The

expressions for these operators are

$$SWAP = \begin{pmatrix} 1 & 0 & 0 & 0 \\ 0 & 0 & 1 & 0 \\ 0 & 1 & 0 & 0 \\ 0 & 0 & 0 & 1 \end{pmatrix}, CNOT = \begin{pmatrix} 1 & 0 & 0 & 0 \\ 0 & 1 & 0 & 0 \\ 0 & 0 & 0 & 1 \\ 0 & 0 & 1. & 0 \end{pmatrix} \quad (22)$$

and

$$T = \begin{pmatrix} i & 0 & 0 & 0 \\ 0 & 1.2389i & 0 & 0 \\ 0 & 0 & 0.8071 & 0 \\ 0 & 0 & 0 & 1 \end{pmatrix} = T_1 \otimes T_2 = \begin{pmatrix} i & 0 \\ 0 & 0.8071 \end{pmatrix} \otimes \begin{pmatrix} 1 & 0 \\ 0 & 1.2389 \end{pmatrix}. \quad (23)$$

It is found that the difference between the operator $C_N$ and $T \cdot SWAP \cdot CNOT$ is the coefficient -1 in the second row of corresponding matrices. Such difference can be eliminated by the phase plate. The operator $T$ can be obtained with $T_1$ and $T_2$. Since the operator $T_1$ can be decomposed as $T_1 = \begin{pmatrix} i & 0 \\ 0 & 1 \end{pmatrix}\begin{pmatrix} 1 & 0 \\ 0 & 0.8071 \end{pmatrix}$, the combination of QWP with the angle $\frac{\pi}{2}$ and PPBS D can realize this operator $T_1$. The operator $T_2$ can be expressed as $T_2 = 1.2389\begin{pmatrix} 0.8071 & 0 \\ 0 & 1 \end{pmatrix}$, which can be implemented by the sandwich type HWP-PPBS D-HWP in the experiment.

**Data availability.** Any related experimental background information not mentioned in the text and other findings of this study are available from the corresponding author upon reasonable request.

**Acknowledgements**

This work was supported by the National key R & D Program of China under Grant No. 2022YFA1404900 and the National Natural Science Foundation of China (12234004 and 12374323).

**Author contributions**

Z. T. and T. C. given the corresponding theory and designed the experiments, Z. T. and X. T performed the experiments, X.D. Z. initiated and designed this research project.

## Competing interests

The authors declare no competing interests.